\documentclass[aps,twocolumn,showpacs,preprintnumbers,amsmath,amssymb,
]{revtex4}

\usepackage{graphicx}
\usepackage{dcolumn}
\usepackage{bm}

\usepackage{color}

\newcommand{\be}[1]{\begin{equation}\label{eq:#1}}
\newcommand{\ee}{\end{equation}}
\newcommand{\bea}{\begin{eqnarray}}
\newcommand{\eea}{\end{eqnarray}}

\newcommand{\bt}{\textbf}

\newcommand{\ph}{\phantom{.}}

\newcommand{\noi}{\noindent}
\newcommand{\no}{\nonumber}

\newenvironment{changemargin}[2]{\begin{list}{}{  \setlength{\topsep}{0pt}  \setlength{\leftmargin}{#1}  \setlength{\rightmargin}{#2}
\setlength{\listparindent}{\parindent}
\setlength{\itemindent}{\parindent}  \setlength{\parsep}{\parskip}
}\item[]}{\end{list}}

\newcommand{\lcm}{\begin{changemargin}{-1.8cm}{0.25cm}}
\newcommand{\ecm}{\end{changemargin}}
\newcommand{\tbcm}{\begin{changemargin}{-0.8cm}{-0.25cm}}
\newcommand{\tecm}{\end{changemargin}}

\begin{document}
\def\v#1{{\bf #1}}

%\preprint{APS/123-QED}

\title{Holographic Charge Density Waves}

\author{Alexandros Aperis$^{*}$}
\author{Panagiotis Kotetes$^{*,\dag}$}
\author{Eleftherios Papantonopoulos$^{*}$}
\author{George Siopsis$^{\flat}$}
\author{Petros Skamagoulis$^{*}$}
\author{Georgios Varelogiannis$^{*}$}
%\email{pkotetes@central.ntua.gr}
%\email{pskam@central.ntua.gr}
\affiliation{$^{*}$Department of Physics, National Technical University of Athens, GR-15780 Athens, Greece\\
$^{\flat}$Department of Physics and Astronomy, The University of Tennessee, Knoxville, TN 37996 - 1200, USA\\
$^{\dag}$ Institut f\"{u}r Theoretische Festk\"{o}rperphysik, Karlsruhe Institute of Technology, 76128 Karlsruhe, Germany}

\begin{abstract}
We discuss a gravity dual of a charge density wave consisting of a
$U(1)$ gauge field and two scalar fields in the background of a
Schwarzschild-AdS$_4$ black hole together with an antisymmetric
field (probe limit). Interactions drive the system to a phase
transition below a critical temperature. We numerically compute
the ground states characterized by modulated solutions for the
gauge potential corresponding to a \textit{dynamically generated
unidirectional charge density wave} in the conformal field theory.
Signatures of the holographic density waves are retrieved by
studying the dynamical response to an external electric field. We
find that this novel holographic state shares many common features
with the standard condensed matter version of charge density wave
systems.
\end{abstract}

 \maketitle

%11.25.Tq    Gauge/string duality
%04.70.Bw    Classical black holes
%71.45.Lr    Charge-density-wave systems (see also 75.30.Fv Spin-density waves)
%71.27.+a    Strongly correlated electron systems; heavy fermions

The AdS/CFT correspondence has become a powerful tool in studying
strongly coupled phenomena in quantum field theory using results from
a weakly coupled gravity background. According to this correspondence
principle \cite{Maldacena}, a string theory on asymptotically AdS
spacetime is dual to a conformal field theory on the boundary.
Recently, following a more phenomenological approach, a plethora of
condensed matter phenomena have been stu\-died within this holographic
principle, like the properties of superfluids and superconductors
both conventional \cite{HartnollPRL101} and unconventional
\cite{Gubser:2008wv}, Fermi liquid behaviour \cite{Cubrovic:2009ye},
non-linear hydronamics \cite{Bhattacharyya:2008jc}, quantum phase
transitions \cite{Iqbal:2010eh} and transport \cite{Herzog:2007ij}.
Thus, classical gravity theories have been transformed into a
laboratory for exploring condensed matter physics from a different
perspective \cite{Koutsoumbas:2009pa}.

The most elucidated example of the application of the AdS/CFT
correspondence to condensed matter physics is the holographic
superconductor. The gravity dual of a superconductor consists of a
system with a black hole and a charged scalar field coupled to a
Maxwell field. The coupling induces a negative effective mass for the
scalar field leading to an instability \cite{GubserInstab}, which
allows to the black hole to have scalar hair at temperatures smaller
than a critical temperature \cite{HartnollPRL101}. In the boundary
dual theory this corresponds to a phase transition at a finite
chemical potential, where the operator dual to the charged scalar
field condenses. Considering fluctuations of the vector potential and
calculating the dynamical conductivity, are crucial for unveiling the
properties of the condensed state.

Apart from superconductivity, understanding real condensed matter
systems demands the introduction of additional ordered states and
their holographic interpretation. The most important of them are
undoubtedly charge and spin density waves (CDW and SDW)
\cite{Gruner}. The development of these states corresponds to the
spontaneous modulation of the electronic charge and spin density,
below a critical temperature ($T_c$). Density waves are widely spread
in different classes of materials and one may distinguish among them,
either orbitally \cite{FISDW} or Zeeman driven \cite{aperis}
field-induced CDWs, confined \cite{GV} and even unconventional
density waves \cite{UnconventionalDW}. The latter play a special role
in condensed matter physics. Characteristic representatives are the
d-density wave (DDW) \cite{DDW} and most recently its chiral (d+id
DDW) version \cite{chiralDDW}, both of which have been proposed as
prominent candidates for explaining the still unidentified pseudogap
phase of the hole-doped cuprates \cite{DDW,chiralDDW} and certain
`hidden' orders in heavy fermion compounds \cite{chiralDDW,heavy}.
Moreover, density waves are also significant due to their strong
tendency to compete or coexist with superconductivity
\cite{Gabovich,CeCoIn5}.

In this Letter, we put forward a gravity dual of a  conventional
charge density wave. Our theory consists of the modulus and phase of
a complex scalar field, a Maxwell field and an antisymmetric field.
We nume\-ri\-cal\-ly obtain ground states characterized by modulated
solutions for the scalar potential corresponding to a CDW in the
boundary dual theory. The solution for the scalar hair corresponds to
a condensed dual operator which follows the usual
Bardeen-Cooper-Schrieffer (BCS) dependence in the vicinity of
$T_{c}$. Signatures of the holographic CDWs are retrieved by studying
the collective modes and the dynamical response to an external
electric perturbation. We find, in the dual boundary theory, a gapped
phason-polariton collective mode \cite{Gruner,Littlewood} with no
Fr\"{o}lich supercurrent \cite{LRA}, which clearly shows that our CDW
is intrinsically commensurate and non-sliding, a novel property
stemming from the gravity-imposed pinning of the scalar potential to
the phason.

Our starting point for constructing the holographic CDW, is the
following Lagrangian density in 3+1 dimensions
\bea
{\cal L}&=&\frac{1}{16\pi G} \left[ R + \frac{6}{L^2} - \frac{1}{6} H^2 \right] \nonumber\\
&-& \frac{1}{4}F^2-\frac{1}{2}\left(\partial_\mu\Phi\right)^2-\frac{1}{2}m^2\Phi^2
-\frac{\lambda_1^2}{2}\Phi^2\left(\partial_{\mu}\vartheta\right)^2 \nonumber\\
&-&\frac{1}{2} \lambda_2^2\Phi^2(B-d\omega^{(2)})(A-d\omega^{(1)})
d\vartheta,\,\label{eq:ExactLagrangian}\eea

\noindent where $A_\mu$ is the Maxwell gauge field of strength
$F=dA$, $B_{\mu\nu}$ is an antisymmetric field of strength $H=dB$ and
$\omega^{(1)}, \omega_\mu^{(2)} $ are auxiliary St\"uckelberg fields.
The last term is a topological interaction term (independent of the
me\-tric). The presence of the St\"uckelberg fields is needed for the
above Lagrangian to be gauge invariant as it can be easily checked if
the following
 gauge
transformations $ A \to A + d \chi^{(1)} $, $\omega^{(1)}\to
\omega^{(1)} + \chi^{(1)}$, $ B \to B + d \chi^{(2)} $ and $
\omega^{(2)} \to \omega^{(2)} + \chi^{(2)}   $ are applied to Eq.
(\ref{eq:ExactLagrangian}). We shall fix the gauge by choosing
$\omega^{(1)} = 0$, $ \omega^{(2)} = 0 $. Apart from the above gauge
symmetries, our model is characterized by an additional global $U(1)$
symmetry, $\vartheta\to\vartheta+\alpha$, that corresponds to the
translational symmetry. This global $U(1)$ symmetry will be
spontaneously broken for $T<T_c$ in the bulk. The concomitant
restoration of this global symmetry will give rise to the related
Nambu-Goldstone mode, the phason \cite{Gruner} as it is called in
condensed matter physics. One may alternatively understand this
global symmetry, by unifying the fields $\Phi$ and $\vartheta$, into
a complex field as $\Psi=\Phi e^{i\lambda_1\vartheta}$, where
$\lambda_1$ corresponds to the charge of the $U(1)$ translational
invariance. In a usual condensed matter CDW, when translational
symmetry is completely broken, the phason is massless and the CDW is
called incommensurate or sliding. The sliding originates from the
freely propaga\-ting phason that gives rise to the Fr\"{o}lich
supercurrent. In spite of the dissipationless electric charge
conduction, $U(1)$ gauge invariance is intact and no Meissner effect
arises. On the other hand, if the phason is gapped then there is a
remnant discrete translation symmetry that prevents sliding and
supresses the Fr\"{o}lich conduction. In this case, the CDW is termed
commensurate or pinned.

The above model is strongly motivated by the usual condensed matter
picture, where a CDW corresponds to the creation of electron-hole
pairs \cite{Gruner}. As a consequence, the complex scalar order
parameter field that describes the transition carries zero electric
charge in stark contrast to the superconducting case where it is
charged. Therefore, the $U(1)$ gauge invariance of the electric
charge will be preserved even in the condensed CDW state. Moreover, the operator that condenses preserves particle number. Therefore, we shall be working with a canonical ensemble minimizing the free energy $F(T,V,N)$ rather than a grand canonical ensemble as is the case with a superconductor which involves a particle number changing order parameter.
The chemical potential will be dynamically generated (unlike with superconductors where one obtains dynamics for a fixed chemical potential).

Despite the
phenomenological character of our theory, that follows closely the
spirit of Ref. \cite{HartnollPRL101}, it is well known that there
exist consistent truncations of M2 string theory resulting in a
supergravity sector with scalar and pseudoscalar fields which are
neutral under $U(1)$ gauge fields \cite{m2}. Such a truncation,
although not consistent, was used in \cite{impurities} to calculate,
using the AdS/CFT correspondence, the momentum relaxation time in a
model where the translational invariance was spontaneously broken due
to impurities.

By varying the metric $g_{\mu\nu}$, we obtain the Einstein equations,
\be{Ein} R_{\mu\nu} - \frac{1}{2} R g_{\mu\nu} - \frac{3}{L^2} g_{\mu\nu} = 8\pi G T_{\mu\nu} \ee
By varying the other fields, $A_\mu$, $B_{\mu\nu}$, $\Phi$ and $\vartheta$, we obtain the field equations, respectively,
\bea\label{eqfield}
\nabla_\mu F^{\mu\nu} &=& \frac{1}{2} \lambda_2^2 J^\nu  \nonumber\\
\nabla_\mu H^{\mu\nu\rho} &=& 8\pi G \lambda_2^2 \mathcal{J}^{\nu\rho} \nonumber\\
\Box \Phi - m^2 \Phi &=& \lambda_1^2 (\partial_\mu\vartheta)^2 \Phi + \lambda_2^2 \frac{1}{\Phi} A_\mu J^\mu \nonumber\\
\nabla^\mu (\Phi^2 \partial_\mu\vartheta) &=& \frac{\lambda_2^2}{2\lambda_1^2} \nabla_\mu K^\mu\eea
where
\bea\label{eqJK} J^\mu &=& \Phi^2 \varepsilon^{\mu\nu\rho\sigma} B_{\nu\rho} \partial_\sigma \vartheta \nonumber \\
\mathcal{J}^{\mu\nu} &=& \Phi^2 \varepsilon^{\mu\nu\rho\sigma} A_\rho \partial_\sigma\vartheta \nonumber \\
K^\mu &=& \Phi^2 \varepsilon^{\mu\nu\rho\sigma} B_{\nu\rho} A_\sigma
\eea
To set the boundary conditions for the various fields, write the metric near the boundary of the asymptotically AdS space in terms of Poincar\'e coordinates as
\be{meas} ds^2 \approx \frac{dr^2}{r^2} + r^2 (-dt^2 + dx^2 + dy^2) \ee
where we set $L=1$.

Suppose that $A_t \equiv V$ and $B^{xy}=-B^{yx}$, $B^{ry}=-B^{yr}$ are the only non-vanishing components of $A_\mu$ and $B^{\mu\nu}$, respectively. We shall also concentrate on static solutions. In the asymptotic region (\ref{eq:meas}), the field equations (\ref{eqfield}) become
\bea
V'' + \frac{2}{r} V' &\approx& 0 \nonumber\\
{B^{xy}}' + \frac{4}{r} B^{xy} - \frac{1}{r^4} \partial_x B^{ry} &\approx& 0 \nonumber\\
\Phi'' + \frac{4}{r} \Phi' - \frac{m^2}{r^2} \Phi &\approx& 0 \nonumber\\
\vartheta'' + 2\left[ \frac{2}{r} + \frac{\Phi'}{\Phi} \right] \vartheta' &\approx& 0
\eea
where we have used the notation $\partial_r \equiv '$. The solutions are
\[ V = V^{(0)} + \frac{V^{(1)}}{r} + \dots \]
\[ B^{ry} = \mathcal{B}^{(0)} + \mathcal{O} (r^{-2}) \ , \ \ B^{xy} = \frac{\partial_x\mathcal{B}^{(0)}}{r^3} + \frac{\mathcal{B}^{(4)}}{r^4} + \mathcal{O} (r^{-5}) \]
\be{assols} \Phi = \frac{\Phi^{(\pm)}}{r^{\Delta_\pm}} + \dots \ , \ \
\vartheta = \vartheta^{(0)} + \mathcal{O} (r^{-1}) \ee
where $\Delta_\pm = \frac{3}{2} \pm \sqrt{\frac{9}{4} + m^2}$. The constant term in the expansion is the source of the corresponding field.
Notice that if the source of the $B^{\mu\nu}$ field ($\mathcal{B}^{(0)}$) is spatially dependent, there is an operator in the CFT dual to $B^{xy}$ that condenses. Moreover, in $B^{xy}$, the coefficient of $r^{-4}$ is arbitrary and represents the vacuum expectation value of a dual operator in the CFT. The latter is similar to the metric giving rise to the stress energy tensor of the boundary CFT. In what follows, we shall concentrate on the simple case in which both $\partial_x \mathcal{B}^{(0)}$ and $\mathcal{B}^{(4)}$ vanish.

We wish to solve the field equations in the {\em probe limit} which is obtained by letting Newton's constant $G\to 0$. In this limit, the Einstein equations (\ref{eq:Ein}) and the equations for the $B^{\mu\nu}$ field (\ref{eqfield}) decouple.
The latter become
$\nabla_{\rho}H^{\rho\mu\nu}=0$. These equations can be solved by fixing
any two-dimensional plane, therefore, by choosing the solution
\be{s1}
B^{ry} = - B^{yr} = 1 \ee
with all other components vanishing,  we
shall  fix the $(r,y)$ plane. This also allows us to set $B^{xy} = 0$ consistently in this limit on account of (\ref{eq:assols}).
We stress here that the decoupling of
the $B^{\mu\nu}$ equation in the probe limit accommodates a constant source for the field (\ref{eq:s1}) allowing the development
of a (1+1)-dimensional CDW on the boundary as we shall see in the
following.

With the above choice of $B^{\mu\nu}$, there is no backreaction to the metric and the Einstein equations (\ref{eq:Ein}) admit the
planar Schwarzschild AdS black hole
\be{s2} ds^2= - f(r) dt^2 +
\frac{dr^2}{f(r)} + r^2 (dx^2 + dy^2) \ \ , \ \ \ \ f(r)=r^2 -
\frac{r_h^3}{r} \ee
as a solution ($L=1$). The metric (\ref{eq:s2}) asymptotes to the AdS metric (\ref{eq:meas}) as $r\to\infty$, as expected.

Having obtained the metric and the $B^{\mu\nu}$ field, the other field
equations come from the Lagrangian
\bea {\cal L}' &=&-\frac{1}{4} F_{\mu\nu}F^{\mu\nu}-\frac{1}{2}\partial_\mu\Phi\partial^\mu \Phi - \frac{1}{2} m^2 \Phi^2\nonumber \\
&&-\frac{1}{2}\lambda_1^2\Phi^2\partial_\mu\vartheta\partial^\mu\vartheta-\lambda_2^2\Phi^2\varepsilon_{ry\rho\sigma}
A^\rho \partial^\sigma \vartheta\,.
\label{eq:ProbeLimitLagrangian}\eea
The resulting field equations are independent of $G$ and
therefore well-defined in the probe limit $G\to 0$. In fact, in
this limit, the last term is of the chiral anomaly type in a $t-x$
spacetime, with the fields depending parametrically on the radial
coordinate. This Lagrangian also appears in the low energy effective
action of the usual (1+1)-D CDWs \cite{ChiralAnomaly}, providing
further support to the phenomenological origin of our theory.

The direction of the modulation is defined by the non zero
components of $B^{\mu\nu}$. Therefore, the Lagrangian appearing in Eq.\
(\ref{eq:ProbeLimitLagrangian}) shows  that the direction of the
modulation is parallel to the $x-$axis and it can be used to
generate any possible profile of modulated charge density  in the
conformal field theory. Our aim is to show that the scalar
potential $A_t \equiv V(r,x)$ asymptotically is of the form
$V(r,x)\sim \mu(x)-\frac{\rho(x)}{r}+\ldots$, with
$\mu(x)\sim\cos(kx)$ and $\rho(x)\sim\cos(kx)$. According to the
AdS/CFT dictionary these two terms of this expansion will
correspond in the dual boundary theory to a chemical potential
$\mu(x)$ and a charge density $\rho(x)$, which are modulated.
Consequently, all the fields will be $x-$dependent.
The chemical potential will be dynamically generated. Thus, we shall be working with a canonical ensemble of fixed particle number. This is possible in our case because the operator that condenses does not change the particle number.

Varying the
Lagrangian (\ref{eq:ProbeLimitLagrangian}) we obtain the field
equations
%%%%%%%%%%%%%%%%%%%%%%%%%%%%%%%%%%%%%%%%%%%%%%%%%%%%%%%%%%%%%
\begin{align}
 \Phi''&+
\left(\frac{f'}{f}+\frac{2}{r}\right)\Phi'+\frac{1}{r^{2}f}\partial_{x}^{2}\Phi-\bigg{\{}\frac{m^{2}}{f}\nonumber\\
&+\lambda_{1}^{2}\left[\left(\vartheta'\right)^{2}+\frac{1}{r^{2}f}
\left(\partial_{x}\vartheta\right)^{2}\right]
+\frac{2\lambda_{2}^{2}}{f^{2}}V\partial_{x}\vartheta\bigg{\}}\Phi=0~,\nonumber\\
\vartheta''
&+\left[\frac{f'}{f}+2\left(\frac{1}{r}+\frac{\Phi'}{\Phi}\right)\right]
\vartheta' +\frac{1}{r^{2}f}\partial_{x}^{2}\vartheta+ \frac{2
\left(\partial_{x}\Phi\right)\left(\partial_{x}\vartheta\right)}{r^{2}f\Phi}\nonumber\\
& + \frac{\lambda_{2}^{2}}{\lambda_{1}^{2}}\frac{2}{f^{2}\Phi}V\partial_{x}\Phi+
\frac{\lambda_{2}^{2}}{\lambda_{1}^{2}}\frac{1}{f^{2}}\partial_{x}V=0~,\nonumber\\
V''&+\frac{2}{r}V'+\frac{1}{r^{2}f}\partial_{x}^{2}V+\lambda_{2}^{2}
\frac{\Phi^{2}}{f} \partial_{x}\vartheta=0~,\end{align}
%%%%%%%%%%%%%%%%%%%%%%%%%%%%%%%%%%%%%%%%%%%%%
%where we have used the notation $\partial_r\equiv\phantom{}'$.
which are also obtained from (\ref{eqfield}) with the background choices (\ref{eq:s1}) and (\ref{eq:s2}).
In the first field equation there is a direct coupling of the scalar
potential $V$ with the phase $\vartheta$. This has important
consequences as we will discuss in the following.

Suppose that the $x$-direction has length $L_{x}$. Then assuming
periodic boundary conditions, the minimum wavevector is \be{s3} k
= \frac{2\pi}{L_{x}}~. \ee Taking Fourier transforms, the form of
the field equations suggests that  it is consistent to truncate
the fields by including $(2n+1)k$-modes for $\vartheta$ and $V$
and $2nk$ modes for $\Phi$, where $n \in \mathbb{Z}$. Thus \bea
\vartheta(r, x) &=& \sum_{n=0}^{\infty}
\vartheta_{(2n+1)k}(r)e^{i(2n+1)kx} + c.c. \nonumber\\
V (r, x) &=& \sum_{n=0}^{\infty} V_{(2n+1)k}(r)e^{i(2n+1)kx} + c.c. \nonumber \\
\Phi(r,
x)&=&\frac{1}{2}\sum_{n=0}^{\infty}\Phi_{2nk}(r)e^{2inkx}+c.c.\label{Fourierfields}
 \eea
We shall consider only the contributions of the leading ($n = 0$)
terms $\vartheta_{\pm k}$, $V_{\pm k}$ and $\Phi_0$. Higher overtones
can be included successively. At each step, we need to add equations
for the new overtones which also introduce corrections to the lower
Fourier modes already calculated in the previous step. Evidently, at
each step the complexity of the system of coupled equations for the
overtones of the three fields increases. We plan on carrying out this
systematic procedure elsewhere.
 Ignoring overtones, the field equations read
\begin{align}
\Phi_0''&+\left(\frac{f'}{f}+\frac{2}{r}\right)\Phi_0'-\bigg\{\frac{m^2}{f}+
2\lambda_1^2\left(\vartheta'_{k}\vartheta'_{-k}
+\frac{k^2\vartheta_{k}\vartheta_{-k}}{r^2f}\right)\nonumber\\
&-2ik\lambda_2^2\frac{V_{k}\vartheta_{-k} -
V_{-k}\vartheta_k}{f^2}\bigg\}\Phi_0=0\,,\label{eq:motion1}\,\\
 \vartheta''_{\pm
k}&+\left[\frac{f'}{f}+2\left(\frac{1}{r}+\frac{\Phi_0'}{\Phi_0}\right)\right]\vartheta'_{\pm
k}-\frac{k^2}{r^2f}\vartheta_{\pm
k}\nonumber\\
&\pm\frac{ik\lambda_2^2V_{\pm k}}{\lambda_1^2f^2}=0\,,
\label{eq:motion2}\\
V''_{\pm k}&+\frac{2}{r}V'_{\pm k}-\frac{k^2}{r^2f}V_{\pm
k}\pm\frac{i\lambda_2^2 k\vartheta_{\pm k} \Phi_0^2}{f}=0\,.
\label{eq:motion3}\end{align}
We shall solve the above equations in the case where the mass of the
scalar field is $m^2=-2$, which is above the Breitenlohner-Freedman
bound \cite{Breitenlohner:1982jf}. We shall also fix the parameters
$\lambda_1=$$\lambda_2 =1$. In order to find the boundary conditions
at the horizon we take the near horizon limit $r\rightarrow r_h$,
where $f(r)\approx f'(r_h)(r-r_h)$, and extracting the parts that
become singular in the above equations we find that $V_{\pm
k}(r_h)=\vartheta_{\pm k}(r_h)=0$, $\Phi_0'(r_h)=m^2\Phi_0(r_h)/3r_h$
and $\vartheta'_{\pm k}(r_h)=\mp ikV'_{\pm k}(r_h)/(3r_h)^2$. At
the boundary, we have asymptotically $\Phi_0 (r)=\frac{\langle {\cal
O}_\Delta\rangle}{r^\Delta}$, where $\Delta = 1,2$ in general, on account of (\ref{eq:assols}). In
this work we focus on $\Delta = 1$ leaving the other case to a future
study.

The fields $\vartheta$ and $V$ are also normalizable and behave asymptotically as in (\ref{eq:assols}) with $\Delta =1$. The coefficients in their respective expansions have a sinusoidal dependence on $x$. Thus we obtain
in the dual boundary theory a
single-mode CDW with a dynamically generated charge density of
the form $\rho(x) = V^{(1)} \sim\cos(kx)$. This case is important because a
lot of materials exhibit either a single-mode CDW or only one
dominant wave-vector. Moreover, the resulting model is numerically
tractable and can provide direct insight on the phase transition
and the response of this novel holographic state.

%Now, let us be more specific and illustrate the exact asymptotic
%behaviour of the fields $\vartheta$, $V$. The general solutions of
%Eqs. (\ref{asympthetaeq}), (\ref{asympVeq}) are
% \bea
% V_{k}(z)&=&\pm\langle {\cal O}_1 \rangle\big[a\cos(uz)+b\sin(uz)\big]\,,\\
% \vartheta_{k}(z)&=&\pm \big[a\cos(uz)+b\sin(uz)\big]\,,\eea
%and
% \bea
% V_{-k}(z)&=&\pm\langle {\cal O}_1 \rangle\big[a\cos(uz)+b\sin(uz)\big]\,,\\
% \vartheta_{-k}(z)&=&\mp \big[a\cos(uz)+b\sin(uz)\big]\,,\eea
%where $u^2 = -k^2 + ik \langle {\cal O}_1 \rangle$, $a$ and $b$ are
%arbitrary constants, and the signs concern the pairs $V_k,
%\vartheta_k$ and $V_{-k}, \vartheta_{-k}$ sepa\-rately. However,
%notice from the above equations that if
%$\vartheta_{k}=\vartheta_{-k}$ then necessarily it should be
%$V_{k}=-V_{-k}$ and if $\vartheta_{k}=-\vartheta_{-k}$ then
%$V_{k}=V_{-k}$ and vice versa. Without loss of generality we consider
%the latter case. Then, substituting the solutions for $\vartheta_{\pm
%k}$ and $V_{\pm k}$ in Eqs. (\ref{Fourierfields}) for $n=0$,
%expanding the $\cos(uz)$ and $\sin(uz)$, since $z\rightarrow 0$, and
%transforming back to the $r$ coordinate we find that asymptotically
% \bea
% \vartheta(r,x)&=& \pm 2i\left[a+\frac{bu}{r}+\cdots\right]\sin(kx)\,,\label{thasympt}\\
% V(r,x)&=& \pm 2 \langle {\cal O}_1
% \rangle\left[a+\frac{bu}{r}+\cdots\right]\cos(kx)\label{Vasympt}\,.
% \eea
%Similar arguments can be used for the solutions with $u^2 = -k^2 -
%ik \langle {\cal O}_1 \rangle$ and retrieve similar results.

\begin{figure}[t]\centering
\includegraphics[width=0.4\textwidth,height=0.30\textwidth]{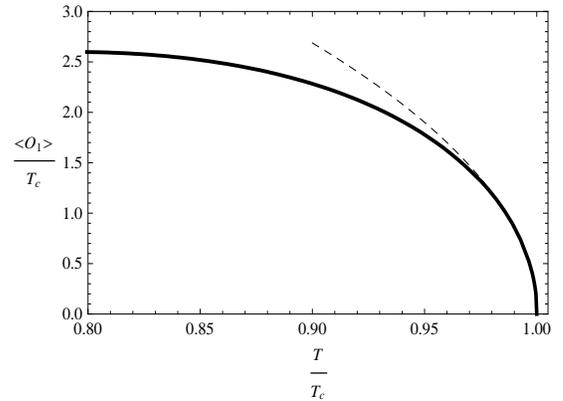}%\vspace{0.45in}%
\caption{Temperature dependence of the condensate (solid line).
%Both axes are normalized with $T_c$.
The dashed line is the BCS fit to the numerical values near $T_c$. We
find $\langle {\cal O}_1\rangle\approx
8.5T_c\left(1-T/T_c\right)^{1/2}$ near $T\rightarrow T_c$.}
\label{fig:GapVsT}
\end{figure}

Having determined the behaviour of the fields both at the horizon
and the boundary, we solve the five field equations
(\ref{eq:motion1})-(\ref{eq:motion3}) numerically setting $k=1$.
We clearly obtain a second-order phase transition with a modulated
scalar potential. Notice that a spatially modulated phase
transition has been also observed in a five-dimensional Maxwell
theory with a Chern-Simons term \cite{nakamura}. In
Fig.\ref{fig:GapVsT} we plot the tempe\-ra\-tu\-re dependence of
the condensate $\langle {\cal O}_1 \rangle$ near the cri\-ti\-cal
temperature. By fitting with $|T-T_c|^{\beta}$, we find that the
critical exponent for the transition is $0.5$, i.e. of the BCS
type.

The critical temperature is controlled by the value of the
wave-vector $k$. Its dependence on $k$ can be deduced from the
scaling symmetry  of the system under the transformation $t\to
\zeta t$, $x\to \zeta x$, $y\to \zeta y$, $r\to r/\zeta$, $V\to
V/\zeta$. It follows that $k/r_h$ is scale-invariant, therefore
\be{s4} T_c\propto k \ee We have checked the validity of this
conclusion numerically. Eq.~(\ref{eq:s4}) should be contrasted
with the case of a holographic superconductor in which $T_c\propto
\sqrt{\rho}$.

Also, notice
that when the condensate $\langle {\cal O}_1 \rangle$ is zero,
i.e. for temperatures above the critical temperature, the
modulated chemical potential and the charge density vanish, and
that they become non zero as soon as the condensate becomes non
zero, i.e. when the temperature is lowered below $T_c$. Therefore,
the modulated chemical potential and the charge density are
\textit{spontaneously ge\-ne\-ra\-ted} and do not constitute fixed
parameters of controlling $T_c$, contrary to what happens in
holographic superconductors \cite{HartnollPRL101} with their
homogeneous analogs $\mu,\rho$. We also have to remark that the
numerical solution of these five equations becomes unstable at low
temperature.
In fact, the fields diverge for $T/T_c\lesssim 0.4$.
This numerical instability indicates that
% the values of
% $\lambda_1$ and $\lambda_2$ are not optimum for our truncated
% model and that a multi-$k$ CDW state may be more favourable for
% $T/T_c\lesssim0.4$. A further analysis is certainly needed,
the probe limit breaks down at low temperatures and
backreaction effects
on the bulk metric become significant.
A further analysis away from the probe limit is certainly needed.

We proceed with examining the collective excitations and response in
the dual boundary theory by considering fluctuations of the
electromagnetic and phason fields of the form
$A_{x}(r,x)$$\rightarrow$ $\widetilde{A}_{\omega,q}(r)e^{i(qx-\omega
t)}$,
$V(r,x)$$\rightarrow$$V(r,x)+\widetilde{V}_{\omega,q}(r)e^{i(qx-\omega
t)}$ and
$\vartheta(r,x)$$\rightarrow$$\vartheta(r,x)+\widetilde{\vartheta}_{\omega,q}(r)e^{i(qx-\omega
t)}$. In the Lorentz gauge, we obtain the equations
\bea\label{eq:flux1}
\widetilde{\vartheta}_{\omega,q}''+\left[\frac{f'}{f}
+2\left(\frac{1}{r}+\frac{\Phi'}{\Phi}\right)\right]\widetilde{\vartheta}_{\omega,q}'
\qquad\qquad\quad\quad\qquad\qquad\no\\
+\left(\frac{\omega^2}{f}-\frac{q^2}{r^2}\right)\frac{\widetilde{\vartheta}_{\omega,q}}{f}
+\frac{i\left(q\widetilde{V}_{\omega,q}+\omega\widetilde{A}_{\omega,q}\right)}{f^2}=0\,,\quad\\
\label{eq:flux2}
\widetilde{V}''_{\omega,q}+\frac{2}{r}\widetilde{V}'_{\omega,q}+\left(\frac{\omega^2}{f}
-\frac{q^2}{r^2}\right)\frac{\widetilde{V}_{\omega,q}}{f}
+\frac{iq\widetilde{\vartheta}_{\omega,q}\Phi^2}{f}=0\,,\quad\quad
%\\
\eea\bea \label{eq:flux3}
\widetilde{A}_{\omega,q}''+\frac{f'}{f}\widetilde{A}_{\omega,q}'+\left(\frac{\omega^2}{f}
-\frac{q^2}{r^2}\right)\frac{\widetilde{A}_{\omega,q}}{f}
-\frac{i\omega(r\Phi)^2\widetilde{\vartheta}_{\omega,q}}{f^2}=0.\no\\
\eea

\noindent In order to get an insight of the response of the our
system to these fluctuations we take the limit $r_h\rightarrow 0$.
%In principle, the dynamics of these fluctuations must be numerically
%determined via solving the above equations. Nevertheless, we may gain
%straightforward insight by considering the zero temperature limit
%$r_h\rightarrow0$. In order to obtain analytical results we consider
%the asymptotic form $\Phi(r)$$\simeq$$<{\cal O}_1>/r$. Since we did
%not obtain a numerical solution for $\Phi(T\rightarrow0)$, we may
%trust this approximation only for inferring qualitative conclusions.
 After setting $z=1/r$ and Fourier transforming $z\rightarrow p$, we obtain
 \bea
\left(\omega^2-q^2-p^2\right)\widetilde{\vartheta}_{\omega,q}(p)
+iq\widetilde{V}_{\omega,q}(p)+i\omega\widetilde{A}_{\omega,q}(p)=0\,,\ph
\label{eq:colq1}\\
+iq\langle {\cal O}_1 \rangle^2\widetilde{\vartheta}_{\omega,q}(p)+\left(\omega^2-q^2-p^2\right)\widetilde{V}_{\omega,q}(p)=0\,,\ph\label{eq:colq2}\\
-i\omega\langle {\cal O}_1
\rangle^2\widetilde{\vartheta}_{\omega,q}(p)+\left(\omega^2-q^2-p^2\right)\widetilde{A}_{\omega,q}(p)=0\,.\ph\label{eq:colq3}
\eea The above homogeneous system of equations defines completely the
dynamics of the fluctuations through simply setting its determinant
equal to zero. Hence, we get
 \bea
\omega^2=q^2+p^2\,,\, \omega^2 = q^2 + p^2 + \omega_g^2 \pm \omega_g
\sqrt{\omega_g^2 + 2p^2}\,,\label{response}\eea

\noi where we have defined $\sqrt{2}\omega_g=\langle{\cal
O}_1\rangle$. Each Fourier $p-$mode of
$\widetilde{\vartheta}_{\omega,q}$, $\widetilde{V}_{\omega,q}$ and
$\widetilde{A}_{\omega,q}$ will give a solution of the form
$e^{i\frac{p}{r}}$, with $p=p(\omega,q)$. Near infinity
$e^{i\frac{p}{r}}\sim 1+i\frac{p}{r},$ and according to the AdS/CFT
correspondence, the first term with $p=0$ corresponds to a source and
the second to a current in the boundary theory. This implies that the
dispersions of the collective modes must be calculated for $p=0$,
while for $p\neq 0$ we obtain the corresponding current that controls
the response of the system such as the conductivity.
\begin{figure}[t]\centering
%\begin{minipage}[b]{3.45in}
\includegraphics[width=0.495\textwidth]{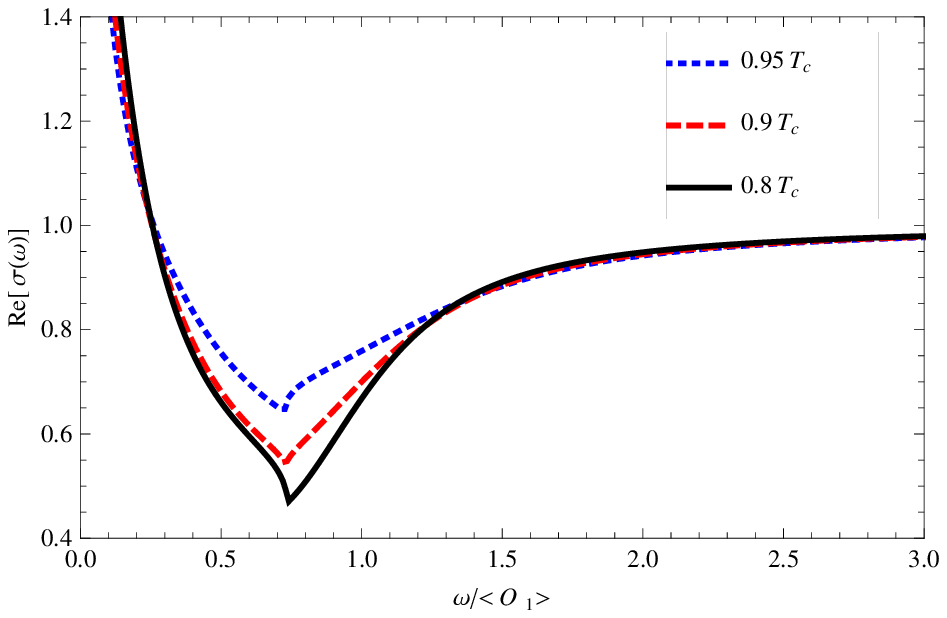}
\hspace{0.00in}
\includegraphics[width=0.485\textwidth]{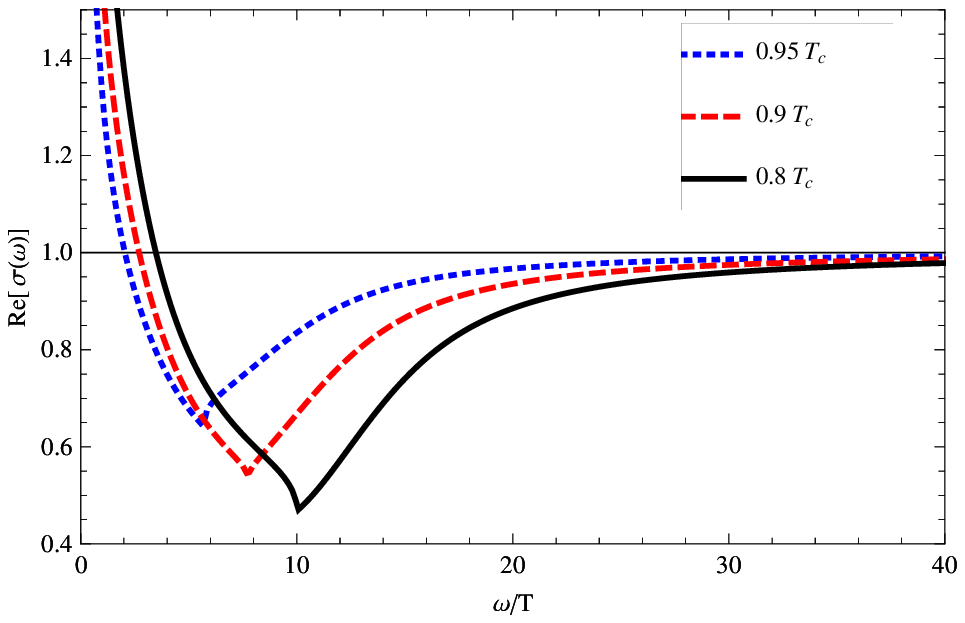}
%\end{minipage}
\caption{The numerically calculated real part of the conductivity
$Re[\sigma(\omega)]$, versus normalized frequency with the condensate
$\omega/\langle {\cal O}_1 \rangle$ (left) and temperature $\omega/T$
(right). In both plots, we clearly observe a `dip' that arises from
the CDW formation and softens with $T\rightarrow T_c$ since $\langle
{\cal O}_1 \rangle\rightarrow 0$. At $T=T_c$ we retrieve the normal
state conductivity $Re[\sigma(\omega)]=1$.}
\label{fig:Conductivities}
\end{figure}

For $p=0$, we obtain from Eq. (\ref{response}) three energy
dispersions $\omega=q$, $\omega=q$, $\omega=\sqrt{\langle {\cal O}_1
\rangle^2+q^2}$. The first two, correspond to massless photonic-like
modes that we anticipated to find since gauge invariance still
persists. On the contrary, the last mode has a mass for $q=0$ equal
to $\langle {\cal O}_1 \rangle$ and basically corresponds to a gapped
phason-like mode which originates from phason-gauge coupling
\cite{Littlewood}. The emergence of the gap hints to the pinning of
the CDW \cite{LRA,ChiralAnomaly}. As a matter of fact, this pinning
has an intrinsic origin, because when the phase transition occurs,
both the scalar potential and the phason field become finite and
modu\-la\-ted by the same wave-vectors $\pm k$. Their existence is
inseparable and in stark contrast to the holographic superconductor
\cite{HartnollPRL101}, the scalar potential is zero above $T_c$. The
interaction term $V(x)\partial_x\vartheta(x)$ in Eq.
(\ref{eq:ProbeLimitLagrangian}), demands that if $V(x)\sim \cos(kx)$
then $\vartheta(x)\sim \sin(kx)$. Their relative phase is `locked' at
$\frac{\pi}{2}$ and the CDW is commensurate.

The dynamical conductivity ($q=0$) is dominated by strongly
hybridized $\widetilde{\vartheta}_{\omega,0}(p)$,
$\widetilde{A}_{\omega,0}(p)$ fluctuations and is defined from
Ohm's law, according to the AdS/CFT correspondence, as
$\sigma(\omega)=\widetilde{A}_{\omega,0}^{(1)}/i\omega\widetilde{A}_{\omega,0}^{(0)}$,
where $\widetilde{A}_{\omega,0}^{(0)}$ and
$\widetilde{A}_{\omega,0}^{(1)}$ are determined from the
asymptotic expansion
$\widetilde{A}_{\omega,0}(r)=\widetilde{A}_{\omega,0}^{(0)}+\frac{\widetilde{A}_{\omega,0}^{(1)}}{r}+\cdots$.
The Fourier expansion
$\widetilde{A}_{\omega,0}(r)=\sum_p\widetilde{A}_{\omega,0}(p)e^{i\frac{p}{r}}$,
implies that asymptotically
$\widetilde{A}_{\omega,0}^{(1)}=i\sum_pp\widetilde{A}_{\omega,0}(p)$
and
$\widetilde{A}_{\omega,0}^{(0)}=\sum_p\widetilde{A}_{\omega,0}(p)$.
Furthermore, we obtain from Eq.~(\ref{response}), for $q=0$ in the
second equation, that
$p_{\pm,\pm}=\pm\sqrt{\omega(\omega\pm\langle {\cal O}_{1}
\rangle)}$, which correspond to two different branches for
propagation direction (first sign). By choosing only the ingoing
contributions $p_{-,\pm}$ we obtain the dynamical
condu\-cti\-vi\-ty
$\sigma(\omega)=\sum_{\epsilon=\pm}\frac{c_{\epsilon}p_{-,\epsilon}}{\omega}$
where $c_{\epsilon}=\widetilde{A}_{\omega,0}(p_{-,
\epsilon})/\widetilde{A}_{\omega,0}^{(0)}$. The factors
$c_{\epsilon}$ will be determined by demanding that
$\lim_{\omega\rightarrow\infty}[\sigma(\omega)-1]\rightarrow0$
faster than $1/\omega$, in order for the Kramers-Kronig relations
to hold. The condition implies $c_{\epsilon}=\frac{1}{2}$ which we
have verified numerically that also holds for finite temperatures.
Kramers-Kronig relations require the fulfillment of the
Ferrell-Glover-Tinkham (FGT) sum rule, dictating that $\int{\rm
d\omega}\ph{\rm Re}[\sigma(\omega)]$ is a constant independent of
temperature. In the super\-con\-ducting case
\cite{HartnollPRL101}, the FGT sum rule demands the pre\-sen\-ce
of $\delta(\omega)$ in ${\rm Re}[\sigma(\omega)]$, giving rise to
a supercurrent. In our case, this rule is sa\-tis\-fied exactly,
without the need of $\delta(\omega)$. The latter reflects the
absence of the Fr\"{o}lich supercurrent, which may be attributed
to the commensurate nature of the CDW.

We now present numerical results for the dynamical conductivity in
the vicinity of $T_c$. In this limit, we are left to solve the system
of the coupled equations Eq. (\ref{eq:flux1}) and (\ref{eq:flux3}).
The stable ingoing boundary conditions at the horizon read
$\widetilde{A}_{\omega,0}(r_h)\propto\sum_{\pm}f^{-i\sqrt{\omega(\omega\pm
r_h\Phi)}/3r_h}$ and
$\widetilde{\vartheta}_{\omega,0}(r_h)\propto\sum_\pm i(\pm
r_h\Phi)^{-1}f^{-i\sqrt{\omega(\omega\pm r_h\Phi)}/3r_h}$. On the
first panel of Fig.\ref{fig:Conductivities} we plot the real
conductivity versus $\omega/\langle {\cal O}_1\rangle$ for different
temperatures, while on the second, the same data are shown versus
$\omega/T$. Notice how the phason-polariton interplay results in a
`dip' structure which gets sharper as temperature decreases.
Moreover, this feature moves to higher frequencies as the condensate
value increases for lower temperatures. Upon raising the
tempe\-ra\-tu\-re, the `dip' CDW fingerprint washes out,
disappea\-ring totally at $T=T_c$ where the normal state is restored.

\textit{In conclusion}, we have proposed a holographic charge density
wave. The dual boundary theory has the characteristic features of a
commensurate CDW, with a gapped phason mode and the absence of a
Fr\"{o}lich supercurrent. The commensurate behaviour due to the
locking of the relative phase of the scalar potential and the phase
field is an intrinsic property of the system and may be attributed to
the presence of the black hole in the bulk. Moreover, the dynamical
conductivity shows phason-polariton mi\-xing. The properties of the
holographic charge density wave certainly need further investigation,
as for example the possibility of a multi-$k$ CDW, the effect of
impurities, the response to an external magnetic field and its
competition to superconductivity. Moreover, it would be interesting
to examine this model away from the probe limit and whether this
phenomenological model can be obtained from a consistent truncation
of string theory.

\textit{Acknowledgements:} We are grateful to A. Kehagias, G.
Koutsoumbas, T. Kolyvaris, H. Soda, M. Tsoukalas, J. Zaanen and V.
Zamarias for stimulating discussions. P.K. acknowledges financial
support from the EU project NanoCTM.  G. S. is supported by the US
Department of Energy under grant DE-FG05-91ER40627. P. S.
acknowledges support from the Operational Program ``Education and
Lifelong Learning'' of the National Strategic Reference Framework
(NSRF) - Research Funding Program: Heracleitus II, co-financed by
the European Union (European Social Fund - ESF) and Greek national
funds.

\end{document}